\documentclass[twocolumn]{aastex62}

\hypersetup{linkcolor=blue,citecolor=blue,filecolor=cyan,urlcolor=magenta}

\usepackage{natbib}
\usepackage{amsmath}

\graphicspath{{./imgs/}}
\received{5.12.2020}
\revised{3.01.2020}
\accepted{5.01.2020}

\shorttitle{A catalog of luminous GC candidates around Cen\,A}
\shortauthors{Voggel et al.}

\begin{document}
\title{A \emph{Gaia}-based catalog of candidate stripped nuclei and luminous globular clusters in the halo of Centaurus\,A}

\correspondingauthor{Karina Voggel}
\email{karina.voggel@astro.unistra.fr}

\author[0000-0001-6215-0950]{Karina T. Voggel}
\affil{University of Utah, James Fletcher Building, 115 1400 E, Salt Lake City, UT 84112, USA}
\affil{Universite de Strasbourg, CNRS, Observatoire astronomique de Strasbourg, UMR 7550, F-67000 Strasbourg, France }
\author[0000-0003-0248-5470]{Anil C. Seth}
\affil{University of Utah, James Fletcher Building, 115 1400 E, Salt Lake City, UT 84112, USA}
\author[0000-0003-4102-380X]{David J. Sand}
\affil{Department of Astronomy/Steward Observatory, 933 North Cherry Avenue, Rm. N204, Tucson, AZ 85721-0065, USA}
\author{Allison Hughes}
\affil{Department of Astronomy/Steward Observatory, 933 North Cherry Avenue, Rm. N204, Tucson, AZ 85721-0065, USA}
\author[0000-0002-1468-9668]{Jay Strader}
\affil{Center for Data Intensive and Time Domain Astronomy, Department of Physics and Astronomy, Michigan State University, East Lansing, MI 48824}
\author[0000-0002-1763-4128]{Denija Crnojevic}
\affil{University of Tampa, 401 West Kennedy Boulevard, Tampa, FL 33606, USA}
\author{Nelson Caldwell}
\affil{Harvard-Smithsonian Center for Astrophysics, Cambridge, MA 02138, USA}




\begin{abstract}

Tidally stripped galaxy nuclei and luminous globular clusters (GCs) are important tracers of the halos and assembly histories of nearby galaxies, but are difficult to reliably identify with typical ground-based imaging data. In this paper we present a new method to find these massive star clusters using \emph{Gaia} DR2, focusing on the massive elliptical galaxy Centaurus\,A (Cen\,A). We show that stripped nuclei and globular clusters are partially resolved by \emph{Gaia} at the distance of Cen\,A, showing characteristic astrometric and photometric signatures. We use this  selection method to produce a list of 632 new candidate luminous clusters in the halo of Cen\,A out to a projected radius of 150 kpc. Adding in broadband photometry and visual examination improves the accuracy of our classification. In a spectroscopic pilot program we have confirmed 5 new luminous clusters, which includes the 7th and 10th most luminous GC in Cen\,A. Three of the newly discovered GCs are further away from Cen\,A in than all previously known GCs.
Several of these are compelling candidates for stripped nuclei. We show that our novel \emph{Gaia} selection method retains at least partial utility out to distances of $\sim 25$ Mpc and hence is a powerful tool for finding and studying star clusters in the sparse outskirts of galaxies in the local universe.

\end{abstract}

\keywords{galaxies: kinematics - galaxies: dwarfs - galaxies: star clusters: general }



\section{Introduction} 
\label{sec:intro}

In hierarchical structure formation, less massive galaxies are accreted by more massive galaxies, and many of the former are torn apart by the tidal forces they encounter \citep[e.g.][]{Wetzel2016}. This tidal disruption process operates
effectively in the low-density outer regions of the accreted galaxies, but the denser central regions
survive and live on in their new parent halos \citep{Bekki2001, Bekki2003, Pfeffer2013, Pfeffer2016}. Most galaxies in the stellar mass range $\sim 10^{8}--10^{10} M_{\odot}$ host dense nuclear star clusters at their centers \citep[e.g.][]{Georgiev2014, denBrok2014, Sanchez2019}. The mass of these nuclear clusters ranges from $\sim 10^6$--
$10^{8} M_{\odot}$, correlating with the stellar mass of the galaxies in which they reside \citep[e.g.][]{Georgiev2016, Sanchez2019}. The nuclear star clusters that survive tidal disruption can therefore be used to trace the assembly history of massive galaxies.

There is compelling evidence that, as a class, many of these stripped nuclei have already been identified amongst the population of ultra-compact dwarfs (UCDs). UCDs are loosely defined as star clusters $\gtrsim 10^6 M_{\odot}$ around galaxies. Most (but not all) UCDs have sizes larger than the typical $\sim$3\,pc half-light radii observed for globular clusters with masses $<$10$^6$~M$_\odot$ \cite[e.g.][]{Norris2014}.

Supermassive black holes were recently found in the centers of five high-mass UCDs \citep{Seth2014, Ahn2017, Ahn2018, Afanasiev2018}. The black holes make up 10-15\% of the total mass of these UCDs, a considerably higher fraction than the typical SMBH in a galaxy. Simulations have shown that such massive black holes ($>10^{6}M_{\odot}$) cannot be made from merged stellar mass black-holes and thus the high mass fraction confirms them as the remnant nuclei of galaxies \citep{Portegies-Zwart2002}.
Other UCDs have shown extended star formation histories unlike those observed in typical globular clusters, which is another piece of evidence that they are stripped nuclei (e.g., \citealt{Norris2015}).

However, while many of the most massive UCDs are likely to be stripped nuclei there is no clear dividing line between globular clusters and stripped nuclei in size or mass. The overall fraction of UCDs that are stripped nuclei, and whether that number changes with mass, is very uncertain \citep{Hilker2006, DaRocha2011, Brodie2011, Norris2011}. A first estimate of the occupation fraction of nuclei among UCDs has been made in \citet{Voggel2019}, based on  integrated dynamical mass estimates that indicate the presence of a measurable central black hole. That paper shows that the majority of UCDs above $M>10^{7}M_{\odot}$ are likely stripped nuclei, with this fraction dropping toward lower masses. Among objects with masses $\lesssim 10^6 M_{\odot}$, only a small fraction have clear signatures of being stripped nuclei, and hence most of these are likely to be globular clusters.

One fundamental challenge in using UCDs to track a massive galaxy's merger history is identifying a complete sample of stripped nuclei. In nearby galaxies, the stellar halos can cover many degrees on the sky, and both stripped nuclei and globular clusters are nearly unresolved from ground-based imaging, making their identification difficult. Even with modern, wide field-of-view spectrographs, it is not currently feasible to obtain spectra of all candidate stripped nuclei. In addition, in some nearby galaxies even radial velocities do not perfectly discriminate between extragalactic objects and foreground Galactic stars.

In this paper we present a method that uses \emph{Gaia}'s exceptional spatial resolution to identify a complete sample of UCDs in Centaurus\,A (Cen\,A; NGC 5128). We chose Cen\,A ($D = 3.8$ Mpc) as the target for this study, as it offers the best place for a feasibly complete search for stripped nuclei in a galaxy that likely hosts many such objects. From previous work, it is clear that Cen\,A hosts a substantial number of luminous UCDs ($L_{\rm V} \gtrsim 10^{6}L_{\odot}$), and that at least a subset of these objects show evidence for being stripped nuclei (e.g., \citealt{Harris2002,Martini2004,Gomez2006,Rejkuba2007}). Deep ground-based imaging of its stellar halo \citep{Crnojevic2016} has revealed that Cen\,A has an active accretion history, with a rich system of satellite galaxies and streams, making it a promising location to search for the remnant nuclei of these stripped galaxies. By contrast, the nearest massive galaxies (the Milky Way and M31)
lack significant UCD populations. UCDs have been well-studied in cluster environments such as Virgo and Fornax, but the larger distances limit detailed follow-up (e.g., adaptive-optics integral field spectroscopy to confirm supermassive black hole) to only the most massive UCDs.
 
While these previous pioneering studies have been crucial to establishing an active accretion history and the presence of UCDs in Cen\,A, 
no existing study of Cen\,A offers a complete sample of UCDs, and most known globular clusters are located within 20kpc of the galaxy center \citep{Rejkuba2007, Beasley2008, Woodley2010, Taylor2016, Taylor2017}. The only work going futher out is \cite{Harris2012} where they catalogued candidate GCs out the 90\,kpc. Yet the halo of Cen\,A extends out to at least 150\,kpc. \emph{Gaia} data, combined with existing wide-field photometry \citep{Taylor2016}, enable us to carry out a nearly complete survey for UCDs in the halo of Cen\,A for the first time.

For the purpose of this work, we use the term UCD to refer to both stripped nuclei and luminous globular clusters; this term has no connotation about the origin of any particular object. Specifically, we use an apparent magnitude cut of $G < 19$
($M_G \lesssim -9.7$; see \S 2.2), equivalent to stellar masses $\gtrsim 10^6 M_{\odot}$. In the Milky Way, the clusters in this mass range include M54, the nucleus of the tidally disrupting Sgr dSph galaxy, and $\omega$~Cen, a likely stripped nucleus \citep[e.g.][]{Bekki2003b}.  Our search also comprises UCDs brighter than these objects that are known to exist in Cen\,A.

We note that while we do not focus on the fainter globular clusters here, these objects are the focus of an upcoming paper (Hughes et al., {\em in prep}).  In that paper, we use the deep high resolution imaging from the Panoramic Imaging Survey of Centaurus and Sculptor (PISCeS) to investigate the population of fainter globular clusters \citep{Crnojevic2014, Crnojevic2016, Crnojevic2019}.  While we share data sources, our approaches differ; here we emphasize on getting a complete sample of bright objects, while in Hughes et al. we focus on building a large sample of globular cluster candidates at large radii.  Many of the sources in our catalog are saturated in the PISCeS imaging, which is why our approach here starts with {\em Gaia}.

Throughout the paper we apply a distance modulus of $m-M = 27.91$ to Cen\,A (\citealt{Harris2010}) and a Milky Way extinction value of $A_{\rm g}=0.379$\,mag \citep{Schlafly2011}.

\section{Characterizing known globular clusters and stripped nuclei in {\it Gaia} and ground-based photometry} 
\label{sec:Gaia}

{\it Gaia} DR2 presents an all-sky astrometric catalog of more than a billion sources \citep{GaiaDR2}. The current effective spatial resolution of the survey (measured by the ability to resolve closely-spaced sources) is $\sim 0.4\arcsec$ \citep{GaiaDR2validation}, though this is expected to improve in future data releases. Since luminous globular clusters and stripped nuclei have effective radii of $\sim$2-10\,pc ($\sim$0.1-0.6\,$\arcsec$ at the distance of Cen\,A), they appear as marginally extended sources to \emph{Gaia}, which has 0.059\,$\arcsec$ pixels in the scanning direction. The principle of our method is that the extended nature of these clusters reveals itself in various astrometric measurements, which can be used to select UCDs in a homogenous manner over the entire 
Cen\,A halo.

\subsection{Astrometric Excess Noise and the BP/RP Excess Factor}
\label{sec:excess}

We have identified two DR2 catalog measurements that are useful for selecting extended sources. The first, which is available for nearly all objects, is the Astrometric Excess Noise (AEN). This  represents the quality of the astrometric fit and is expected to be zero when all observations fit the model of an individual star perfectly \citep{Lindgren2018}. Extended sources have a higher excess noise compared to a point source. 

The second measurement is the BP/RP Excess Factor (hereafter abbreviated as $BR_{\rm excess}$), which is available for most bright (\emph{Gaia} $G < 19$) sources. This is derived from a comparison of the $G$ (broadband) with the Blue Photometer ($BP$, 3300-6800\,\AA) and Red Photometer ($RP$, 6400-10500\,\AA) magnitudes. While the $G$ magnitude is derived from profile fitting with an effective resolution of $\sim 0.4\arcsec$, the $BP$ and $RP$ magnitudes are derived directly from the flux within an aperture of $3.5\times2.1\,\rm arcsec^{2}$ \citep{GaiaDR2validation}. $BR_{\rm excess}$ is defined as the relative flux ratio of the $BP$ + $RP$ fluxes and the $G$ flux. Even for a point source, this value is slightly larger than unity since the BP and RP filters (taken together) have a broader wavelength range than the $G$ filter. For extended sources, the ratio is even larger, since the ``extra" light beyond the point source is picked up in the larger $BP$ + $RP$ apertures. However, a high ratio could also indicate contamination from a nearby source (expected in crowded regions), and extended sources could be background galaxies or double stars rather than star clusters.

Most {\em Gaia} sources in the magnitude range of potential GCs have both excess factors. Off all sources in the vicinity of Cen\,A 98.8\% of sources between G=15-19 mag and 97.3\% of sources in between G=19-20 mag have the two excess factors.

To test whether these quantities can effectively select extended sources in \emph{Gaia} DR2, we use a set of confirmed UCDs (massive globular clusters and candidate stripped nuclei) in Cen\,A to see whether they have elevated AEN and $BR_{\rm excess}$ factors. The test sample of velocity-confirmed clusters is taken from various literature sources \citep{Taylor2017,Woodley2010,Beasley2008}. We require that all sources have existing photometric measurements, and select only those objects with $g_0 < 18.8$, corresponding to an apparent limit of $g < 19.1$. We also exclude sources within 5\,$\arcmin$ ($\sim 5.5$ kpc) of the center of Cen\,A owing to issues with crowding and extinction. The systematic velocity of Cen\,A is $v_{\rm helio}=541$ km\,s$^{-1}$ and the dispersion of the GCs is $\sigma \sim 150$ km\,s$^{-1}$ \citep{Peng2004}. Thus, we use a conservative radial velocity cut of $v < 275$ km s$^{-1}$ to remove objects with ambiguous velocities that might instead be Galactic foreground stars.  This leaves us with a final set of 61 confirmed UCDs as our test sample.

When we match this sample to \emph{Gaia} DR2, we find that all of the UCDs have a \emph{Gaia} match within 1\arcsec (see also \S 2.2). This indicates that the {\em Gaia} catalog has a high completeness for Cen\,A star clusters with $g < 19.1$, at least away from the more crowded central regions of the galaxy.

As a next step, we checked whether any of these literature objects had significant parallaxes or proper motions in {\em Gaia} (with S/N $>$ 4). This would indicate that these ``confirmed" UCDs are actually misclassified foreground stars. We found that four of the 61 object indeed have significant proper motions, indicating that they are Galactic stars. These four objects all had low S/N measurements of their radial velocities in \citet{Taylor2010} and hence it is likely they were misclassified from poor velocity measurements. In Table \ref{tab:foreground} these mis-classified GCs are listed. A complete list of the 57 literature globular clusters used in this comparison can be found in Table \ref{tab:lit_gcs} in the Appendix.

\begin{figure}
   \centering
   \includegraphics[width=\hsize]{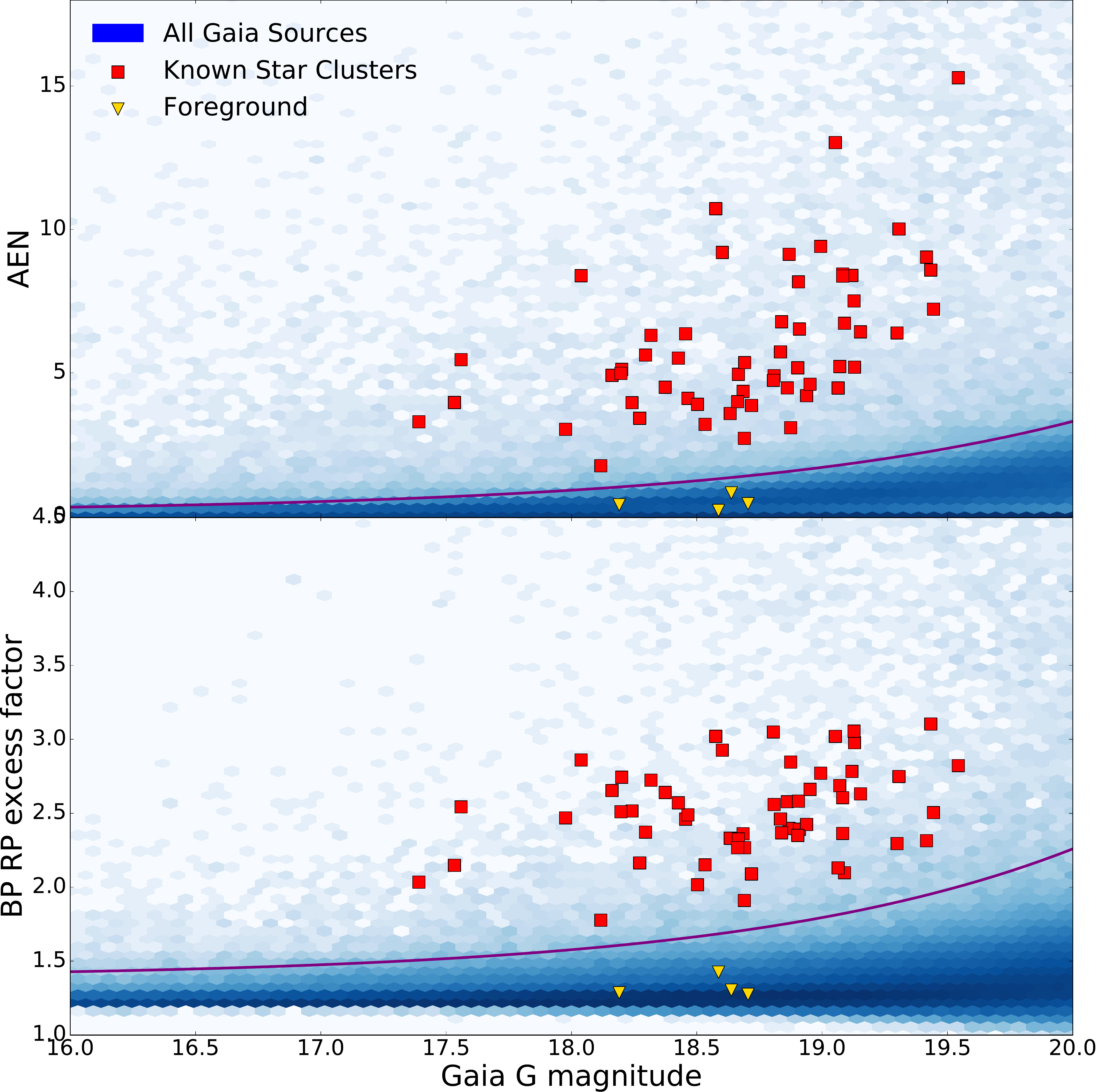}
      \caption{Top: Astrometric Excess Noise (AEN) plotted against {\em Gaia} G magnitude. The blue density plot shows all {\em Gaia} sources within 2.3$^{\circ}$ of Cen\,A. Red squares are confirmed bright ($g<19.1$) literature globular clusters and yellow triangles are misclassified literature globular clusters (actually foreground stars with high significance proper motions). Bottom: As above, but for 
      $BR_{\rm excess}$ rather than AEN. Both panels show that these astrometric excess factors can effectively select extended objects (included stripped nuclei and globular clusters) in Cen\,A .} 
         \label{fig:completeness}
   \end{figure}

The two astrometric quantities (AEN and $BR_{\rm excess}$) are plotted against $G$ for this sample of 61 literature UCDs in Figure \ref{fig:completeness}, using different symbols for the misclassified objects. We compare this sample to that of all \emph{Gaia} sources within 2.3$^{\circ}$ of Cen\,A. The main stellar locus is clearly visible in both panels as a dark blue plume at low values of these excess factors. Notably, the confirmed UCDs all have AEN and $BR_{\rm excess}$ significantly above the stellar locus, showing that these astrometric excess factors can clearly identify the extended nature of these sources.

Guided by these results, we draw boundaries between the extended UCDs and the stellar locus at the 98th percentile of the distributions of AEN and $BR_{\rm excess}$ as a function of magnitude. Cutting sources below the 98th percentile does not remove any literature sources while getting rid of a maximum of stellar contaminants. We then fit the following exponential functions to represent the boundaries:

\begin{equation}
\label{eq:AEN}
   {\rm AEN} = 0.12 + 2.66 \times 10^{-6} e^{0.7 G} 
   \end{equation}
 \vspace{-5mm}
\begin{equation}
\label{eq:BPRP}
   BR_{\rm excess} = 1.39 + 2.18 \times 10^{-7} e^{0.76 G} 
   \end{equation}

Figure \ref{fig:completeness} shows that these boundaries effectively separate confirmed UCDs from stars.

 \begin{deluxetable*}{cccccccc}
\tablecaption{List of foreground stars that were mis-classified as globular clusters in the literature}
\tablehead{\colhead{Name}  & \colhead{Name}  & \colhead{R.A.} & \colhead{Dec.} & \colhead{Radial velocity} & \colhead{Vel. Source} & \colhead{g mag} & \colhead{{\em Gaia} R.A. Proper Motion}  \\
\colhead{Beasley+2008} & \colhead{Taylor+2017} & \colhead{} & \colhead{} & \colhead{km/s} & \colhead{} & \colhead{mag} & \colhead{$\mu$as/yr}  \label{tab:foreground}} 
\startdata
--  &   GC0565 &  201.0275 & -42.882944 &  285.0 $\pm$  29.0 & Woodley+2010  &  17.99 & -9.82 $\pm$ 0.35 \\   
AAT109380 &  GC0047 & 201.181583 & -43.145333 &  448.0  $\pm$  31.5 & Beasley+2008  & 18.48  & -4.66 $\pm$ 0.53 \\
PFF-gc039  &   GC0133  & 201.287917  & -42.40025 &  388.2  $\pm$  25.2  & Beasley+2008  & 18.61 & -3.58 $\pm$ 0.52 \\
PFF-gc046  &  GC0159  & 201.311792 & -43.686278 &  526.6  $\pm$  20.4  & Beasley+2008  &  18.61 & -4.20 $\pm$ 0.65 \\
\enddata
\end{deluxetable*}
 
\subsection{How does the \emph{Gaia} photometric system compare to the literature photometry?}
\label{sec:gaiaphot}

Here we compare \emph{Gaia} photometry to that already published in the literature. In particular, \citet{Taylor2017} presents multi-band photometry for a large number of sources around Cen\,A\footnote{In the course of this work we discovered that the photometry from Table 2 of \citet{Taylor2017}, which states that it is not corrected for foreground extinction, has in fact already had these corrections applied. This photometry is discussed in detail in Hughes et al, in preparation.}.
In the median, the $G-g$ values for our sample of literature UCDs is 0.19, with an rms scatter of 0.16 mag. This modest scatter shows that we have in fact correctly identified \emph{Gaia} sources for the clusters. Note that the magnitudes quoted here are all extinction corrected unless explicitly mentioned; we assume uniform extinction corrections of $A_{\rm V}=0.32$, $A_{\rm G}=0.27$, $A_{\rm g}=0.37$ and $A_{\rm r}=0.26$ \citep{Schlafly2011}. 

At first consideration, the derived $G-g \sim 0.19$  is surprising, since the $G-g$ color is expected to be $\sim -0.4$ to $-0.7$ for the $g-r$ colors of our GCs using the relations from \cite{Jordi2010}. To investigate this, we also match stars from \citet{Taylor2017} to \emph{Gaia} that are at the same $u-r$ colors range as the confirmed GCs. We find an offset between the stellar locus and the globular clusters in $G-g$ of 0.52mag.

This tells us that the measured $G$ magnitudes of GCs in Gaia are half a magnitude \emph{fainter} than expected since the UCDs are resolved and the $G$ mags are measured assuming the profile of a point source. Indeed, this effect is what makes the identification of UCDs with \emph{Gaia} possible. 

Correcting our $G-g$ colors by this 0.52\,mag offset leads to a range of $G-g$ colors of 0.0 to -0.67 with a mean of -0.32.  This erases the bulk of the discrepancy between the observed and expected colors of our sources but the GCs are still an average of 0.2mag too red compared to the theoretical colours of \cite{Jordi2010}.

As another check of the fraction of the flux we are missing from Cen~A globular clusters, we matched the \emph{Gaia} $G$ magnitudes to a sample of Cen\,A globular clusters from \citet{Peng2004} which have carefully-calibrated $V$ mags. We find a
$G-V\sim0.29$, with a dispersion of 0.17 mags. This compares to an expected $G-V$ of $\sim$-0.25 for typical GC colors \citep{Jordi2010}, suggesting an offset of 0.5-0.6 mags in $G$, consistent with the 0.52mag update we found above. 

Using the observed $G-V$ color suggests our $G<19$ limit corresponds to $M_V \lesssim -9.5$ after correcting for extinction.  This corresponds to a $V$ band luminosity of $5\times10^5$~L$_\odot$, which assuming the typical $M/L_V$ of 2 found for globular clusters \citep[e.g.][]{Voggel2019} suggests a cluster mass of 10$^6$~M$_\odot$.  Thus our survey limit corresponds roughly to clusters above 10$^6$~M$_\odot$.  

For the remainder of the paper, whenever we convert the \emph{Gaia} $G$ magnitudes to absolute magnitudes or luminosities, we account for this effect. For instance, the $G<19$ cut used for our sample corresponds to $M_G \sim -9.7$ after correction for extinction {\em and} inclusion of this offset.

\subsection{Correlation between excess factors and size}
\label{sec:excess}

The previous part of this section shows that two \emph{Gaia} astrometric parameters can reliably identify known UCDs in Cen\,A. Here we test the ability to obtain even more information: since the excess factors quantify how much a given source deviates from a single star model, in principle, larger UCDs should also have larger excess factors.

Only a subset of our sample of confirmed Cen\,A UCDs have existing reliable size measurements. We find 21 such objects in the literature \citep{Rejkuba2007, Taylor2010, Woodley2010}.
To enlarge the sample and test the correlation over a larger magnitude range, we also add 19 UCDs with $G < 20$ and measured sizes from the Virgo Cluster study of \citet{Liu2015}.

   \begin{figure*}
   \centering
   \includegraphics[width=\hsize]{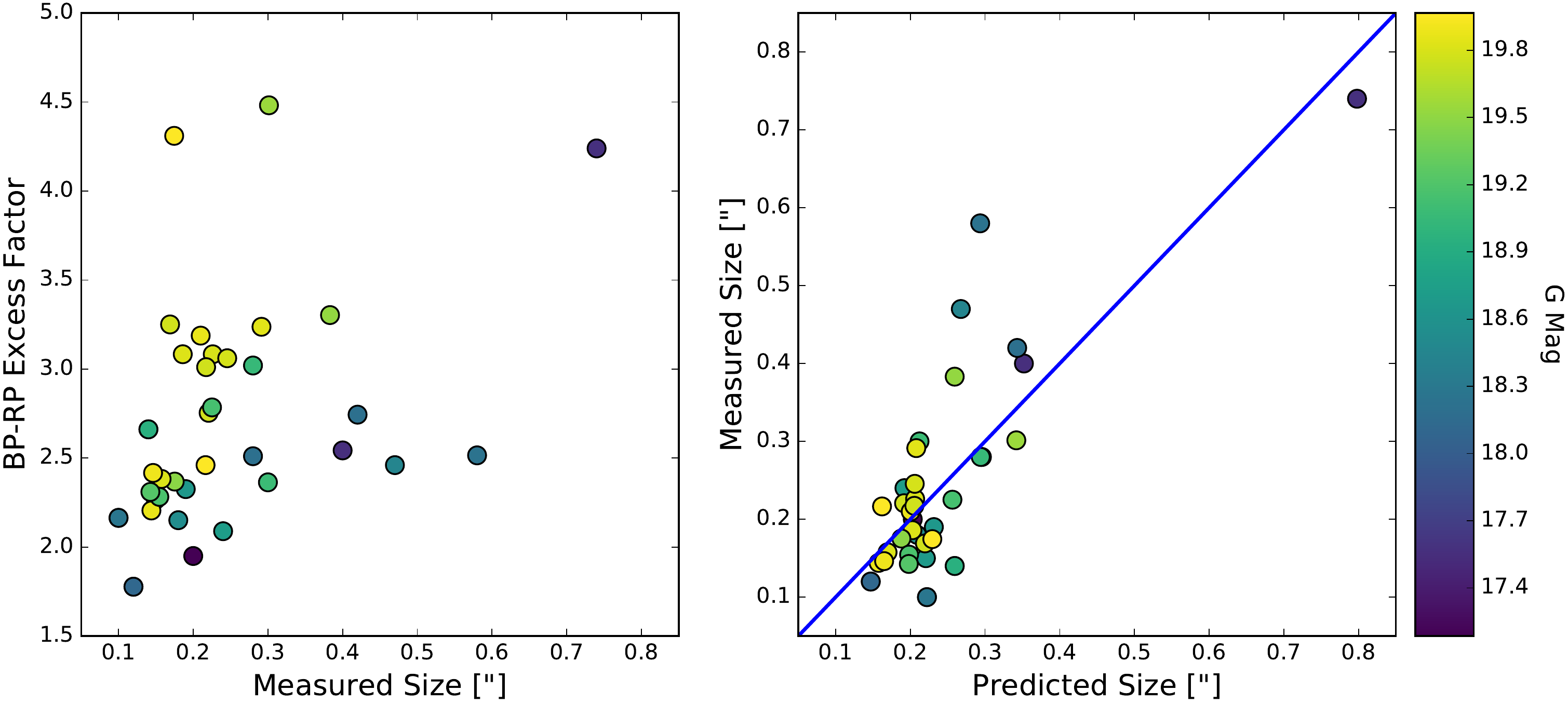}
     \caption{Left: $BR_{\rm excess}$ as a function of the measured sizes of UCDs in Cen\,A and Virgo. The shading scales with the $G$ magnitude of the source. These quantities are somewhat correlated, but with a large magnitude-dependent scatter. Right: The predicted size from 
      equation \ref{relation} compared to the actual size, showing the improved quality of the correlation once magnitude is considered. This relation has a scatter of about 30\%.} 
         \label{fig:excess_sizes}
   \end{figure*}

Figure \ref{fig:excess_sizes} compares the angular sizes of this composite sample of UCDs with $BR_{\rm excess}$. These quantities show a correlation; however, the correlation is magnitude-dependent, since $BR_{\rm excess}$ also depends on $G$
(see Figure \ref{fig:completeness}).

To account for this, we fit a function that depends on both the magnitude and $BR_{\rm excess}$ to the data:

\begin{equation}
 \begin{split}
r_h (") =  (0.222 -0.095(G-18))({BR_{\rm excess}}-2.5)\\ + 0.308(G-18)-0.074
 \end{split}
\label{relation}
\end{equation}

This resulting relation is shown in the right panel of Figure \ref{fig:excess_sizes}. While the range of the predictions is modest and the scatter is substantial (rms of $0.081\arcsec$, which is a fractional scatter of 31\%), it still suggests that it is possible to measures the sizes of UCDs with \emph{Gaia} data to an accuracy of $\sim$30\% over the size range $\sim 0.1$--0.5\arcsec. This accuracy is lower than can be achieved using, for example, the \emph{Hubble Space Telescope}, but it is superior to the quality of ground-based sizes in most cases. It would be worthwhile to extend this work to a larger sample over more uniform range of galaxy distances in the future.

The work above solely concerns $BR_{\rm excess}$: the AEN measurements have lower S/N, especially for fainter sources, so appear less promising than $BR_{\rm excess}$ for this purpose.

\subsection{Broadband colors of star clusters}
\label{sec:mixture_colours}

Another source of information available is $ugriz$ broadband photometry from the Survey of Centaurus A's Baryonic Structures \citep[SCABS;][]{Taylor2016, Taylor2017} program. They published a catalog of $3200$ GC candidates as well as photometry of $\sim500,000$ point sources in their survey area, extending out to $D_{\rm proj}$$\approx$150 kpc from CenA.

It is well-established that the integrated stellar populations of globular clusters separate in color-color space from contaminants including foreground stars and background galaxies, and that the quality of the separation increases with the width of the baseline (e.g., \citealt{Strader2011,Munoz2014,Taylor2017}). Here we use the $u-r$ vs. $r-z$ color space, where globular clusters are well-separated at the metal-rich end, although there is some overlap at the metal-poor end.

   \begin{figure}
   \centering
   \includegraphics[width=\hsize]{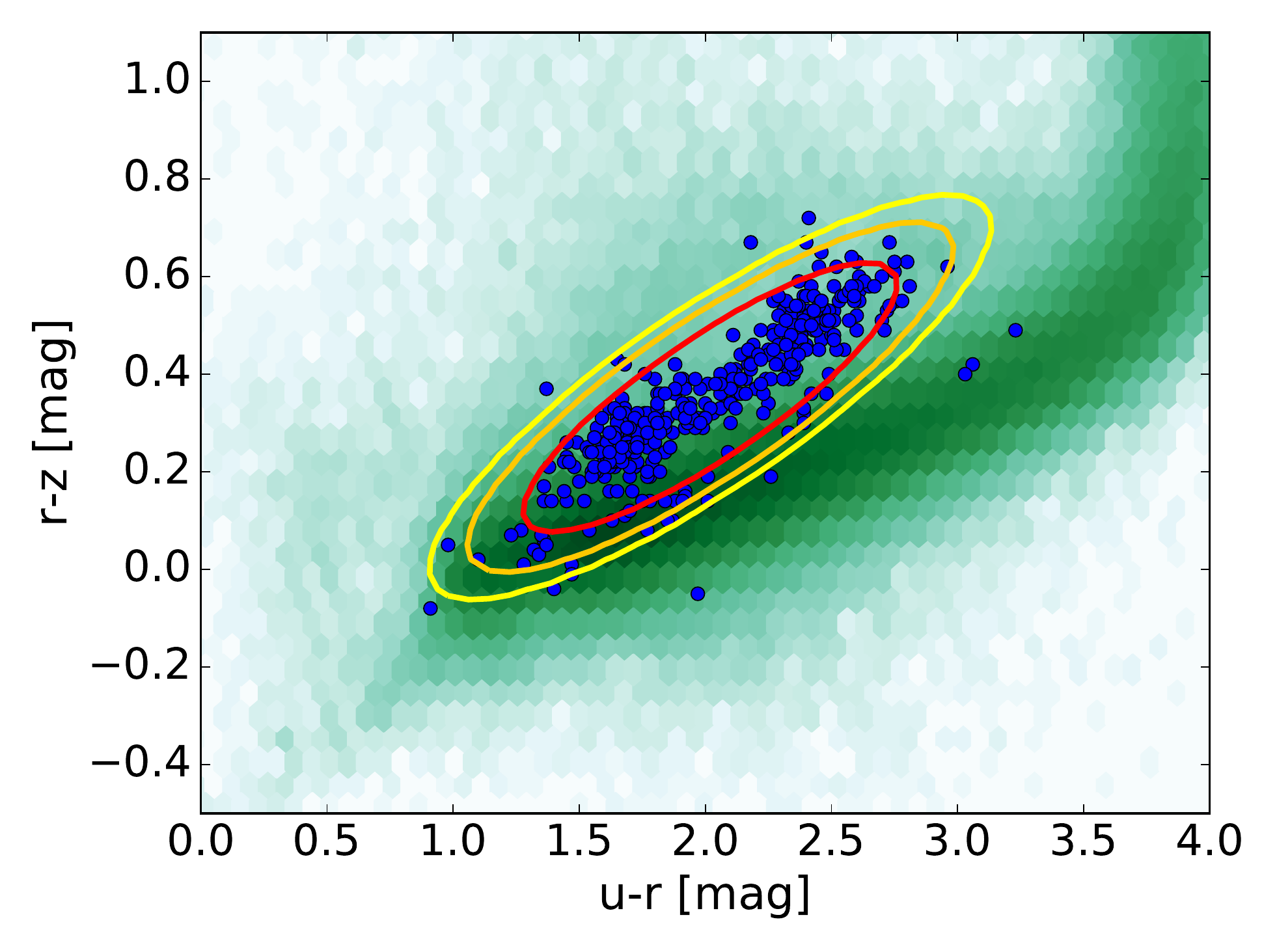}
      \caption{Color--color plot of velocity confirmed star clusters (blue) and stars (green) around Cen A, with all extinction-corrected photometry taken from \citet{Taylor2017}. The red, orange, and yellow ellipses represent the results of mixture modeling that include 85\%, 95\% and 97\% of all confirmed globular clusters.} 
         \label{fig:colour_taylor}
   \end{figure}  

To do this we first matched the position of velocity-confirmed literature globular clusters/UCDs \citep{Harris1992,  Harris2002, Peng2004, Woodley2005, Harris2006,  Rejkuba2007, Woodley2007, Beasley2008,  Woodley2010b, Woodley2010a, Taylor2010} with the \citet{Taylor2017} photometric catalog of point sources, finding 389 matches. In Figure \ref{fig:colour_taylor} we show this color-color plot and, as a comparison, \citet{Taylor2017} colors for all \emph{Gaia} sources with $G < 19$ within 2.3$^{\circ}$ of the center of Cen\,A. The colors use the magnitudes of their Table 2, which are already extinction corrected. As expected, the clusters are well-separated from the dense stellar locus, except at the bluest end of their distribution.

To have a quantitative way to estimate the probability of a photometric source being a UCD based on these color--color data, we use a Gaussian mixture model \citep{scikit-learn} to separate the clusters and stars. The results of this model (see Figure \ref{fig:colour_taylor}) provide a likelihood for each source to be classified as a star cluster.

\section{Finding new luminous globular clusters and stripped nuclei with {\it Gaia}}

Now that we have explored how known globular clusters and stripped nuclei look in the {\it Gaia} excess factors and in broadband colors, we can apply this knowledge to the entire {\it Gaia} catalogue around Cen\,A to find new candidate UCDs.

Our work here is based on the second data release of the {\it Gaia} mission \citep{GaiaDR2, GaiaDR2validation}. We download all the {\it Gaia} DR2 sources within a radius of 2.3$^{\circ}$ of the center of Centaurus\,A (R.A.: 13h25m27.6152s, DEC.: -43d01m08.805s), which is equivalent to a physical radius of 150\,kpc, a radius encompassing the imaging data collected by the PISCeS project \citep{Crnojevic2016}. This leaves us with an initial catalogue of 514,439 {\it Gaia} sources. From here
we make several cuts to narrow down the catalog to the most likely star cluster candidates.

Our selection method follows these steps:
\begin{enumerate}
\item We select all {\it Gaia} sources with $G<19.0$. This corresponds to UCDs with $M_G \lesssim -9.7$, equivalent roughly to masses $\gtrsim 10^6 M_{\odot}$ as discussed in Section~\ref{sec:gaiaphot}.

\item We eliminate foreground stars by cutting sources that have a \emph{Gaia} proper motion in either coordinate or a parallax measurement of S/N = 4 or greater, since Cen A UCDs should not have measurable values of these quantities at \emph{Gaia} precision.

\item We next apply a cut in the Astrometric Excess Noise and $BR_{\rm excess}$. For this we use the exponential cut-off line to the two excess factors described in equations \ref{eq:AEN} and \ref{eq:BPRP} and shown in Figure \ref{fig:completeness}. 
\item Our set of 389 confirmed Cen A star clusters has a color range of 0.66-1.6 in $BP-RP$. Thus, we also apply a broad {\em Gaia} color cut selecting only sources with $0.6 < $BP-RP$ < 1.6$ to remove objects from the extreme red and blue end of the distribution where no known globular clusters are located.

\end{enumerate}

\begin{figure*}
   \centering
   \includegraphics[width=\hsize]{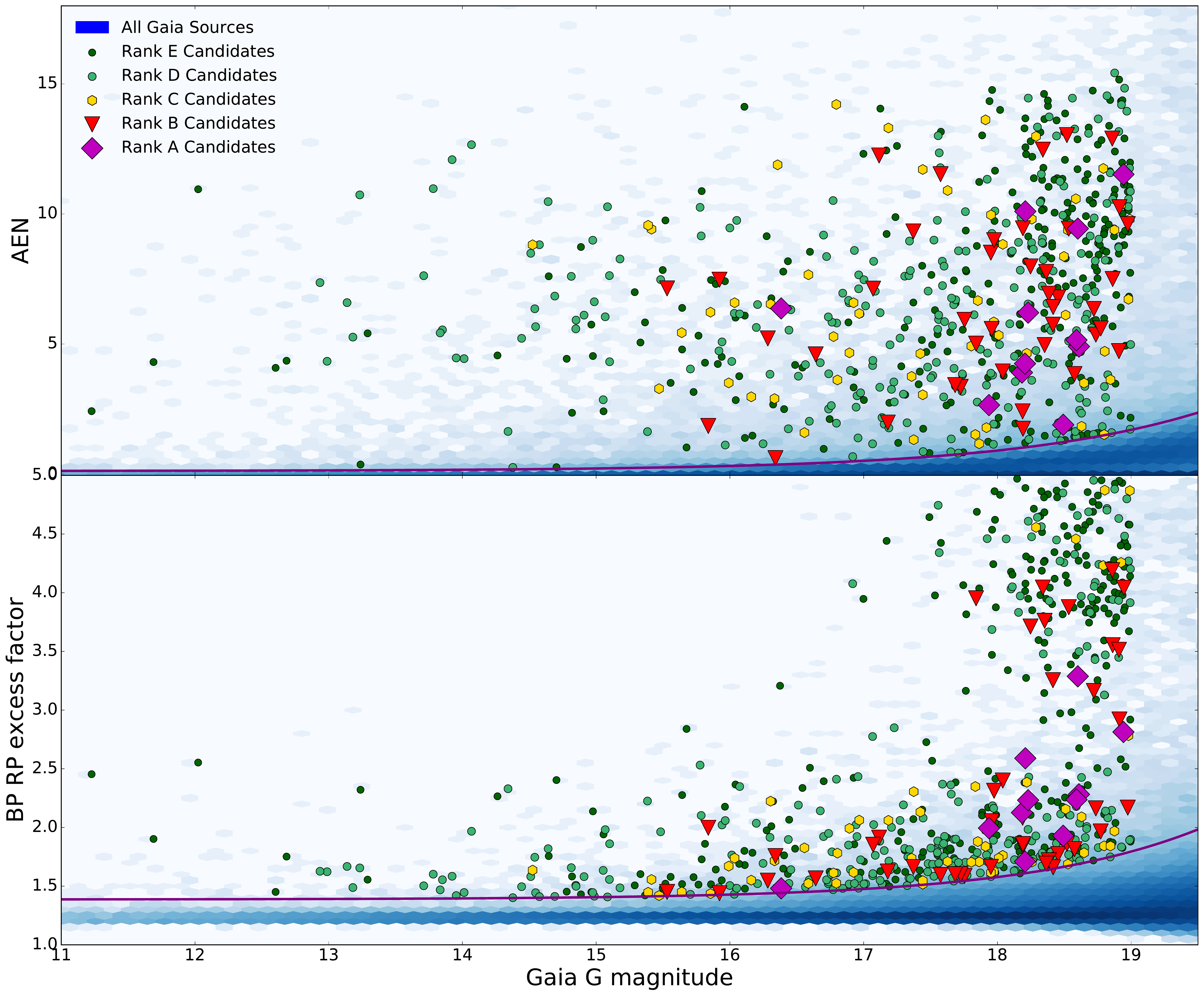}
      \caption{The distribution of the 632 selected {\em Gaia} sources that pass our UCD selection criteria in the two excess factors against $G$ magnitude. The color coding is the final ranking of the candidates (including visual classification), where Rank A candidates (large purple diamonds) are the most likely UCDs and Rank E (small dark green circles) the least likely.} 
         \label{fig:selected_candidates}
   \end{figure*}

With these selection criteria we get a final list of 632 new candidate UCDs. The distribution of these candidates in AEN and $BR_{\rm excess}$ as a function of magnitude is plotted in Figure \ref{fig:selected_candidates}. 
We note that making only a cut on AEN would result in 1530 candidates while making a cut on $BR_{\rm excess}$ only would result in 840 candidates. This suggests both Gaia parameters provide useful information for creating a complete sample while minimizing contamination.

\subsection{Color information for the {\em Gaia} selected cluster candidates}

\begin{figure*}
   \centering
   \includegraphics[width=\hsize]{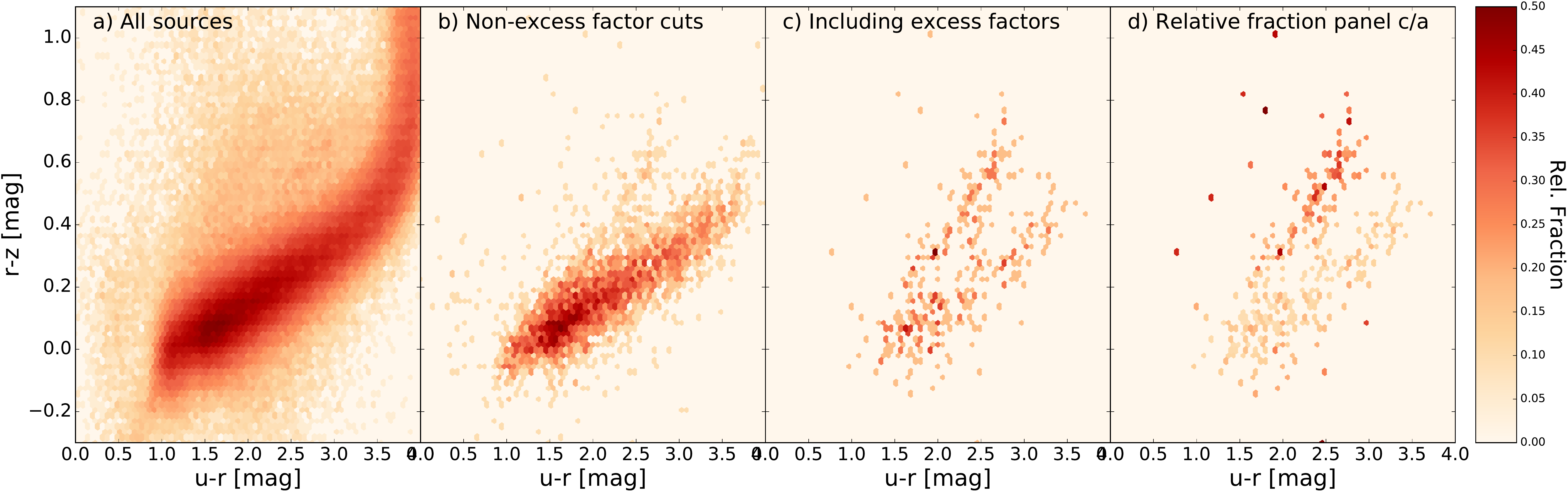}
      \caption{$r-z$ vs.~$u-r$ color--color plots, showing the stages of our UCD candidate selection. The left panels shows the density distribution of all point sources from \citet{Taylor2017} Table 2 that have a {\em Gaia} match. The second panel has our cuts in $G$, $BP-RP$, parallax, and proper motion. The third panel adds cuts in $BR_{\rm excess}$ and AEN. The last panel is the ratio of the third and first panels, i.e., the relative fraction of the third panel with full photometry. This shows that our method does effectively select likely UCD candidates.} 
         \label{fig:colour_gaia}
   \end{figure*}

While we do not yet have the information necessary to definitively determine how many of these candidates are true UCDs, we can investigate our selection criteria function by showing how the density of sources in color--color space changes as we apply our cuts sequentially. The first panel of Figure \ref{fig:colour_gaia} shows the density distribution of all \emph{Gaia} sources that also have multi-band photometry in Table 2 of \citet{Taylor2017}.
The star cluster sequence is not visible in this panel as it is washed out by the scatter in the overwhelming stellar locus.

The next panel shows the effects of the cuts on $G$, $BP-RP$, proper motion, and parallax. This panel is still dominated by a thick stellar locus, but the hint of the UCD sequence is emerging. The third panel finally adds the AEN and $BR_{excess cuts}$, where the sequence of likely Cen~A objects is now strong. The final panel is a ratio of the third and first panels, with the UCD
sequence boldly visible, showing that our criteria effectively 
select UCDs, especially among the redder objects where there is less stellar contamination.  This figure also shows that {\em Gaia} data can also significantly improve on color-only selection methods.

\subsection{Visual classification of candidates}

For our next  classification stage, we used data from the Panoramic Imaging Survey of Centaurus and Sculptor (PISCeS, \citealt{Crnojevic2014,Crnojevic2016}) taken with Megacam on Magellan, which has imaging data for 346 (about $\sim 55\%$) of the 632 UCD candidates; the rest do not fall within the survey footprint. The typical angular sizes of UCDs at the distance of Cen\,A are 0.1--0.6\arcsec ($\sim$2-11\,pc). In good seeing conditions ($\sim$0.5-0.6\arcsec) the outer regions of the UCDs 
start to resolve into individual RGB stars, providing a clear  signature of their identity. Such data also work to reject contaminants: background galaxies and close (``double") sets of foreground stars can pass the {\em Gaia} astrometric cuts but be readily visible by eye.

To formalize our visual inspection of these PISCeS data, we used $1.2'\times1.2'$ cutout images centered on each candidate. After training ourselves on confirmed UCDs, team members carefully examined each image for evidence that the candidate object was 
resolved into stars in its outskirts or more extended than surrounding point sources. Five team members voted independently on the likelihood that each candidate was a star cluster. These five votes were then averaged to generate a final visual assessment score for each target.

\begin{figure*}
   \centering
   \includegraphics[width=\hsize]{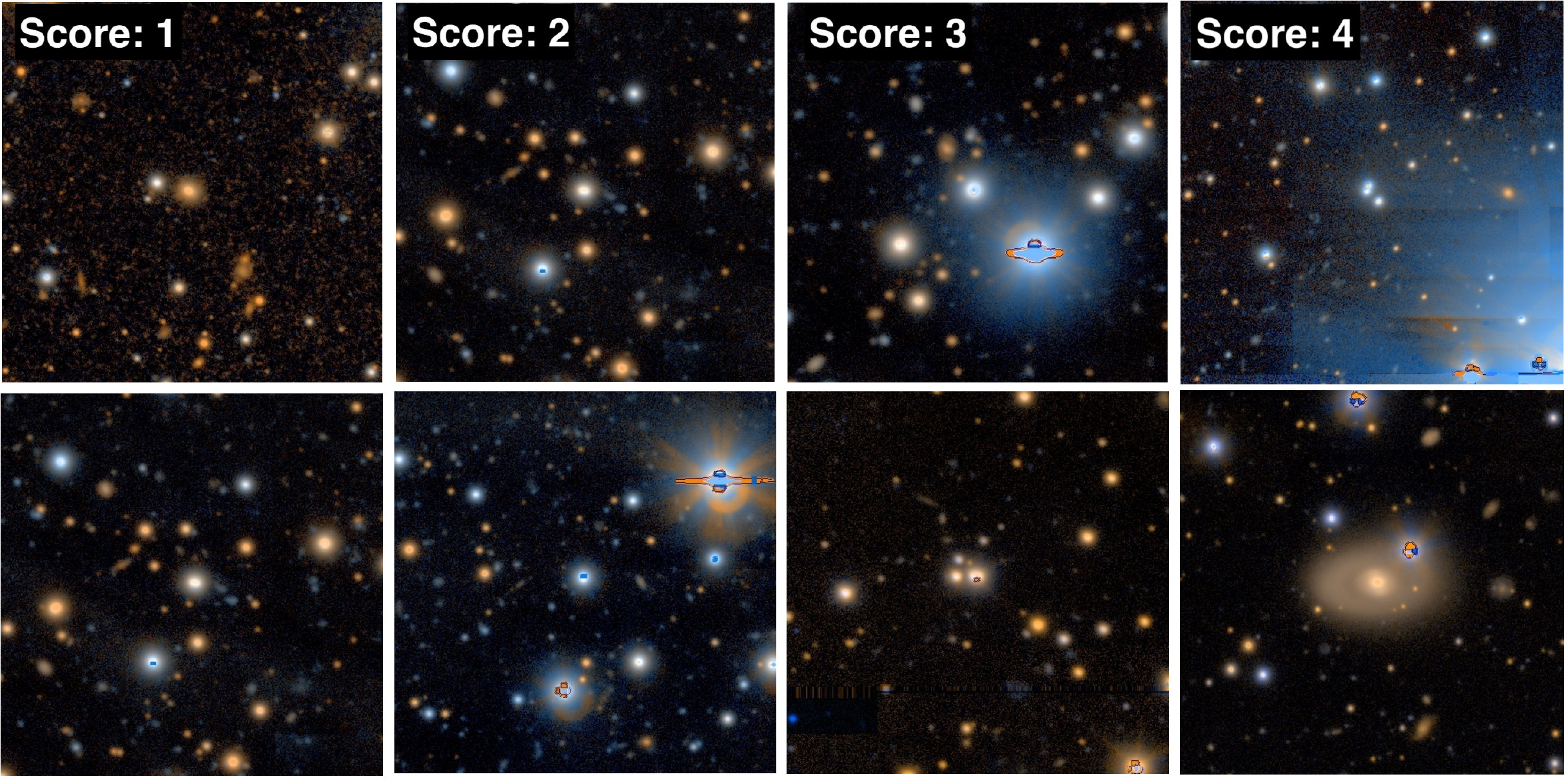}
      \caption{Shown here are $1.2'\times1.2'$ cutouts of PISCES imaging \citep{Crnojevic2016} for two examples in each category of visual assessment score, as described in the text. Each panel is centered on a UCD candidate identified with the Gaia selection criteria. Score 1 objects are the best UCD candidates; Score 2 and 3 are less likely; Score 4 are definite contaminants such as background galaxies (bottom right) or double stars (top right).}
         \label{fig:stamps_categories}
   \end{figure*}

On the basis of a ``pilot program" for this visual grading, we decided on four categories, ranging from 1 (most likely to be a UCD) to 4 (least likely). Category 1 UCDs showed clear signs of a resolved stellar halo. Category 2 UCDs showed hints of being extended but without obviously resolved outskirts. Category 3 objects had no evidence for or against them being UCDs; this category often includes objects taken in poor seeing conditions as well. Category 4 objects are obvious non-clusters: typically double stars or background galaxies. We show two examples from each category in Figure \ref{fig:stamps_categories}.
A histogram showing the distribution of all candidates we voted on is shown in the top right panel of Figure \ref{fig:ranking_summary}.

\subsection{Final ranking of Star cluster candidates}

   \begin{figure*}
   \centering
   \includegraphics[width=\hsize]{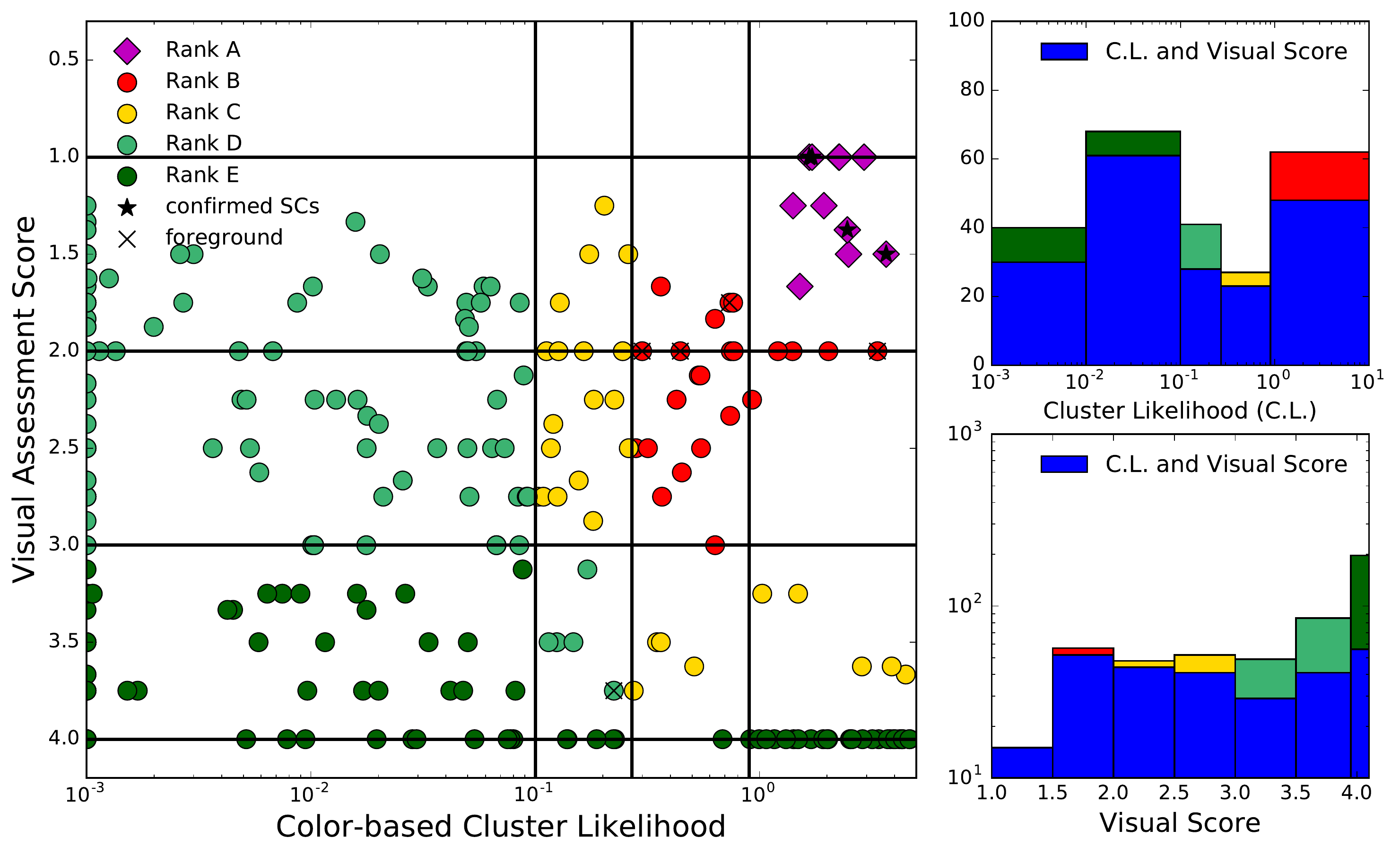}
      \caption{A summary of the ranking of all candidates. The left panel shows targets for which both a color-based likelihood and visual assessment score information are available. Ranks A to E represent most likely to least likely UCDs, with the colors as Rank A (purple); Rank B (red); Rank C (yellow); Rank D (light green); Rank E (dark green). The vertical black lines mark the 0.1, 0.27 and 0.9 relative cluster likelihood levels, which correspond to the three color ellipses (red, orange, and yellow) in Figure \ref{fig:colour_taylor}. Points with a black star in their centers are newly velocity-confirmed UCDs (\S 4), whereas the black crosses mark candidates that are newly confirmed as foreground stars. The histograms on the right show the distribution of all votes and all cluster likelihood values. The blue portion of the histogram marks candidates for which both quantities are available (color and PISCeS imaging), with the added colored bars on top showing the distribution of candidates for which only cluster likelihood or votes are available. The colors of these added bars reflect the associated ranks.}
         \label{fig:ranking_summary}
   \end{figure*}  
   
We refine our initial UCD candidate selection where possible 
by adding in information from (a) the $urz$ color selection 
 discussed in Section~\ref{sec:mixture_colours} and (b) our visual assessment score from PISCeS imaging discussed above. This 
 additional classification information is not available for all 632 \emph{Gaia} UCD candidates. Of the 346 candidates with PISCeS data, 278 have colors and 68 do not. Of the 286 candidates without PISCeS images, 225 have colors and 61 do not.

Here we discuss our methodology for the final ranking of UCD candidates. 

\begin{figure}
   \centering
   \includegraphics[width=\hsize]{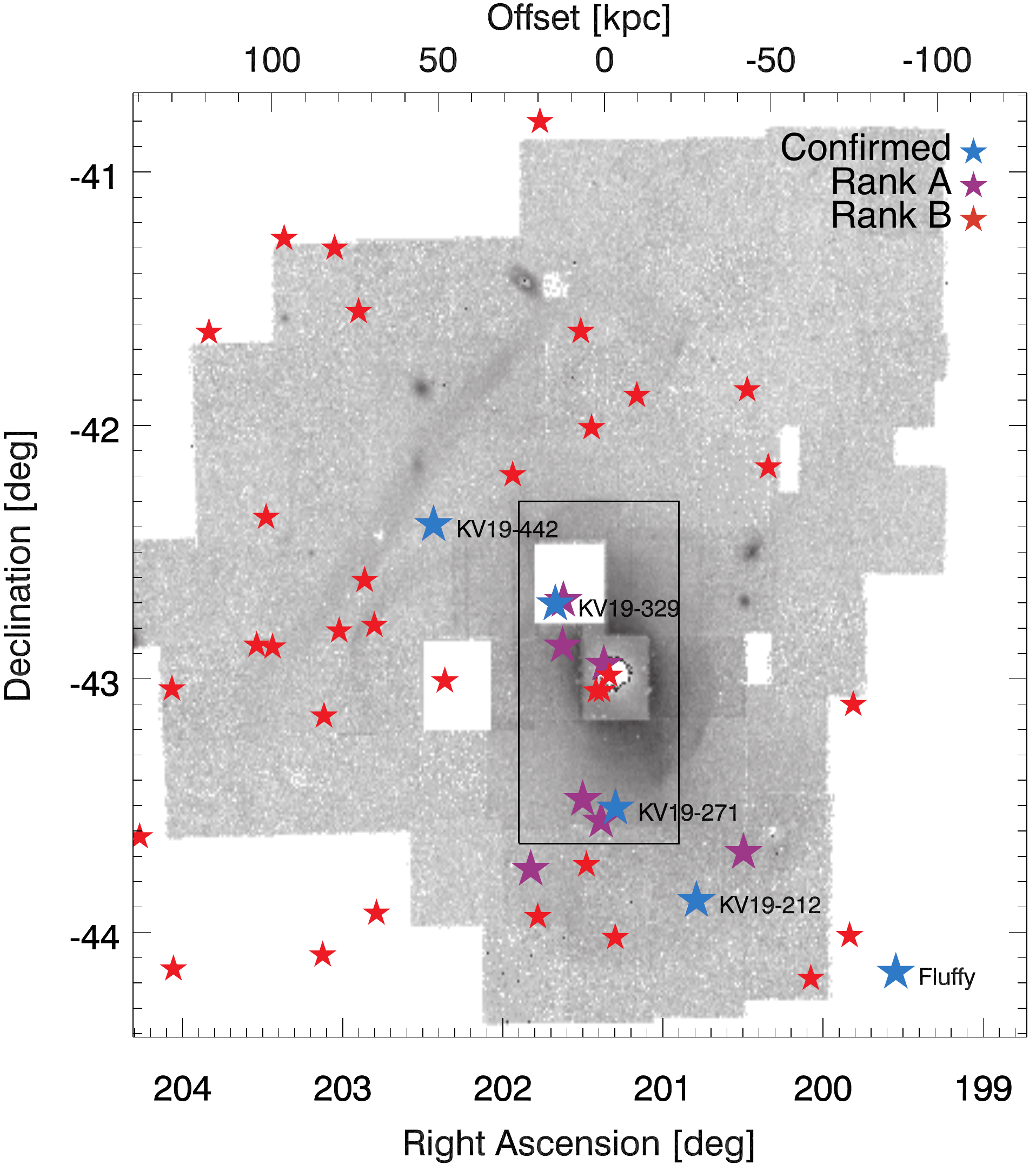}
      \caption{The location of the 11 candidates with Rank A is marked with purple stars, where the 5 new confirmed UCDs (see Table \ref{tab:kinematics}) are shown as blue labeled data points. The Rank B candidates are shown as red stars. The black box marks the area that contains all previously confirmed globular clusters in Cen\,A. The background image depicts the stellar halo of Cen\,A from the PISCES imaging \citep{Crnojevic2016}  } 
         \label{fig:location_map}
   \end{figure}

\vspace{2mm}
\noindent \textbf{Rank A, Most Probable UCDs:}  Our Rank A candidates are the most likely UCDs, based on both their colors and visual inspection. This group includes objects with colors that correspond to cluster likelihoods of $>$0.9 (the inner-most ellipse in Figure~\ref{fig:colour_taylor}), and visual assessment scores $<$2.0 (Figure~\ref{fig:stamps_categories}). Figure \ref{fig:ranking_summary} visually shows their relative ranking in these color and visual assessment categories. Only 11 objects are in this rank, and spectroscopic follow-up of 4 of them (Section \ref{sec:followup}) have confirmed all as UCDs around Cen\,A. Their spatial distribution in the halo of Cen\,A is shown in Figure \ref{fig:location_map}.
 
\vspace{2mm}
\noindent \textbf{Rank B, Good candidate UCDs:}
Rank B candidates were selected in one of two ways. In the first, the visual or color information strongly suggests that the objects are UCDs (visual assessment score $\leq$2.0 or cluster likelihood$\geq$0.9), but only one of these two pieces of information is available. The other selection method for this rank is candidates with moderately favorable scores for both categories 
(visual assessment score $\leq$3.0 and cluster likelihood$\geq$0.27) but that were not high enough for Rank A.
There are 42 Rank B candidates. As for Rank A, the spatial distribution of these candidates around 
Cen\,A is shown in Figure \ref{fig:location_map}. 

\vspace{2mm}
\noindent \textbf{Rank C, Candidate UCDs:}
Rank C includes candidates where the color information and PISCeS imaging data do not suggest a high probability that they are true UCDs, but where this identification is still possible. The details of the classification of objects in this rank can be seen by  examining Figure \ref{fig:ranking_summary}. 54 {\em Gaia} candidates fall into this rank.

\vspace{2mm}
\noindent \textbf{Rank D, Candidate UCDs with no additional evidence or conflicting information:}  Rank D includes the 61 candidates with no PISCeS imaging or color information. It also includes 185 candidates with contradicting information from the votes and probabilities (see Figure \ref{fig:ranking_summary}).

\vspace{2mm}
\noindent \textbf{Rank E, Not UCDs:}
Rank E candidates have colors or imaging that show they are not UCDs. This rank includes all candidates with colors inconsistent with being a UCD and with visual assessment scores $\geq 3.0$, plus all objects with visual assessment scores of 4 (all team members agreeing that the object is a contaminant). There are 279 {\em Gaia} candidates which fall into this rank.

A visual summary of these ranks can be found in Figure \ref{fig:ranking_summary}, and complete information for all 632 candidates, including the \emph{Gaia} data, colors and magnitudes from \citealt{Taylor2017} if available, and other classification information, can be found in Table \ref{tab:all_cands}. The Taylor colors and magnitudes are all extinction corrected.



\subsection{Intrinsic completeness of {\em Gaia} UCD candidates}
One goal of the present work is to provide a complete census of UCDs in the halo of Cen\,A. Here we analyze the completeness of our sample of new candidates.  

In Section \ref{sec:excess}, we showed that all 61 previously known velocity-confirmed UCDs with $g < 19.1$ are detected by {\em Gaia} and pass our astrometric selection, suggesting a high completeness for this \emph{Gaia}-based UCD search. However, we have visually identified a very extended ($r_h \sim 40$\,pc) massive ($\sim 10^6 M_{\odot}$) star cluster (named ``Fluffy"), in the outskirts of Cen\,A using our PISCeS imaging and with spectroscopic follow-up (see next Section; Crnojevic et al., {\em in prep}). This object does not appear in the  {\em Gaia} source catalogue at all, although its integrated magnitude is theoretically bright enough to make our cuts. 

We know from section \ref{sec:gaiaphot} that we  miss 0.5\,mag of extended sources in Gaia. That Fluffy is not in {\em Gaia} is thus not surprising as its large size and lower central surface brightness means that the flux enclosed in the {\em Gaia} aperture may make it too faint to have been cataloged. Diffuse UCDs have been detected in other galaxies,
mostly in the Virgo and Fornax Clusters \citep{Brodie2011,Strader2012,Forbes2013,Voggel2016a}.
They make up a small fraction of the total UCD population, so this source of incompleteness is unlikely to be significant, though such objects are strong candidates to be stripped nuclei and hence are worthy of detailed study. Moreover, we have not identified other objects with similar properties in our visual inspection of PISCeS images. In addition, the completeness of \emph{Gaia} sources drops toward fainter magnitudes.
For example, the completeness for velocity-confirmed Cen\,A globular clusters is $\sim 90\%$ out to 20\arcmin at $G = 20$ (Hughes et al., in prep). For the sources actually in \emph{Gaia}, nearly all have the astrometric measurements we use to identify candidate UCDs.

Overall our source list appears to be highly complete.  Our final candidate list most highly ranks objects where both PISCeS imaging and photometry from \citet{Taylor2017} are available.  This is true for only 278 out of 632 (44\%) candidates, suggesting that more complete imaging and photometry data could roughly double our sample of Rank A candidates.

\begin{deluxetable*}{cccccccccccc}
\tablecaption{List of \emph{Gaia} UCD Candidates. \label{tab:all_cands}}
\tablehead{\colhead{Name}  & \colhead{R.A.}  & \colhead{DEC} & \colhead{AEN} & \colhead{{\em Gaia} G} & \colhead{BP RP Excess} & \colhead{u-r} & \colhead{r-z} & \colhead{r mag}& \colhead{Votes} &  \colhead{C.L.} & \colhead{Rank} \\
\colhead{}  & \colhead{deg}  & \colhead{deg} & \colhead{} & \colhead{mag} & \colhead{} & \colhead{mag} & \colhead{mag} & \colhead{mag}& \colhead{} &  \colhead{} & \colhead{}} 
\startdata
KV19-001 & 198.276663 & -42.993805 & 2.16 & 18.01 & 2.02 & 1.03 & -0.16 & 16.96 & nan & 0.01 & D \\
KV19-002 & 198.322490 & -43.583197 & 3.76 & 16.51 & 2.19 & nan & nan & nan & nan & nan & C \\
KV19-003 & 198.334218 & -43.114592 & 7.02 & 18.73 & 3.43 & nan & nan & nan & nan & nan & C \\
KV19-004 & 198.363842 & -43.735719 & 8.24 & 18.34 & 3.48 & nan & nan & nan & nan & nan & C \\
KV19-005 & 198.364438 & -43.382436 & 8.82 & 14.58 & 1.41 & nan & nan & nan & nan & nan & C \\
\enddata
\end{deluxetable*}

\section{Follow up of UCD candidates}\label{sec:followup}

While all of our highly-ranked UCD candidates will eventually require spectroscopic follow-up to confirm their nature, here we present the results of our initial efforts to canvass these candidates with spectroscopy.

We obtained spectroscopic follow-up of 14 UCD candidates using the Magellan Inamori Kyocera Echelle spectrograph \citep[MIKE;][]{Mikeref} on Clay/Magellan. We observed three UCD candidates (KV19-442, KV19-329 and KV19-271) on the two nights of 16-17 June 2018 and 10 additional UCD candidates on 5-6 April 2019, as well as the
UCD candidate ``Fluffy" that was found from PISCeS rather than \emph{Gaia}. All observations consisted of 2x1800s exposures.

   \begin{figure*}
   \centering
   \includegraphics[width=\hsize]{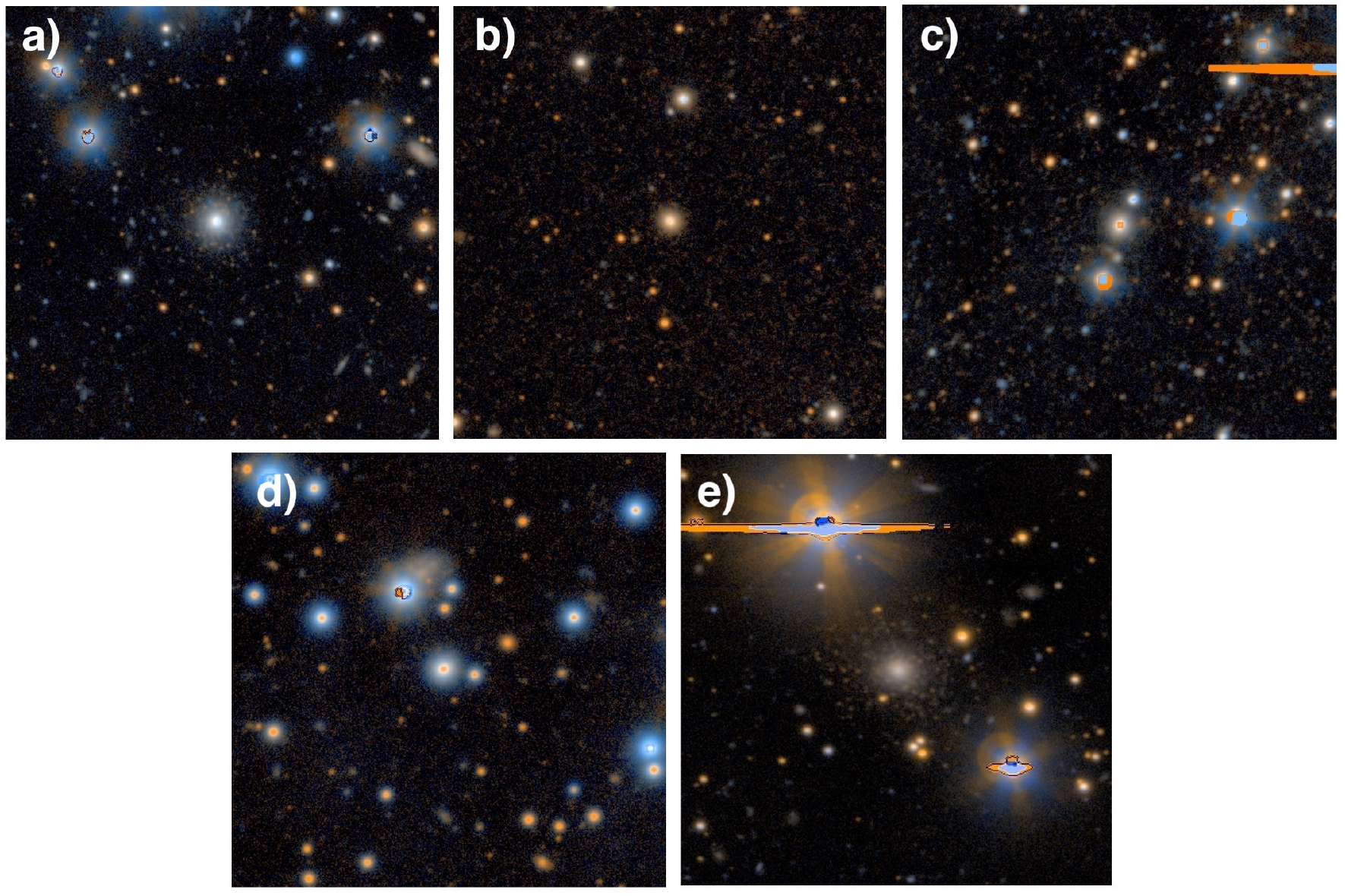}
      \caption{The $1.2' \times1.2'$ cut-outs show the PISCES imaging \citep{Crnojevic2016} of the five confirmed UCDs in Centaurus\,A. The first four panels (a--d) are \emph{Gaia}-selected Rank A UCD candidates, all of which are confirmed; the last panel (e) is the visually-selected source ``Fluffy''.}

         \label{fig:stamps_confirmed}
   \end{figure*}  
   
We reduced the MIKE spectra using the {\tt CarPy} pipeline \citep{Kelson2000, Kelson2003}. We focus on the spectral order that contains the Ca triplet from 8470-8700\,\AA, which are the deepest and most prominent lines of the spectra. We determined the radial velocities by modeling the spectra of the UCD candidates using a set of 10 template stars observed during the same run with the pPXF package \citep{Cappellari2004}. Here we present only the velocities; a future paper will describe the data and analysis in greater detail, including velocity dispersion measurements and dynamical masses (Dumont et al., {\em in prep}).

The systematic velocity of Cen\,A is $v_{\rm helio}=541$ km\,s$^{-1}$ and the dispersion of its GC system is as high as $\sigma \sim 150$ km\,s$^{-1}$ \citep{Peng2004}. 
Stripped nuclei might or might not have a similar velocity dispersion to globular clusters, but in any case, we can use the GC kinematics as a rough guide to values that should be reasonable for the UCD candidates. 

\begin{figure}
   \centering
   \includegraphics[width=\hsize]{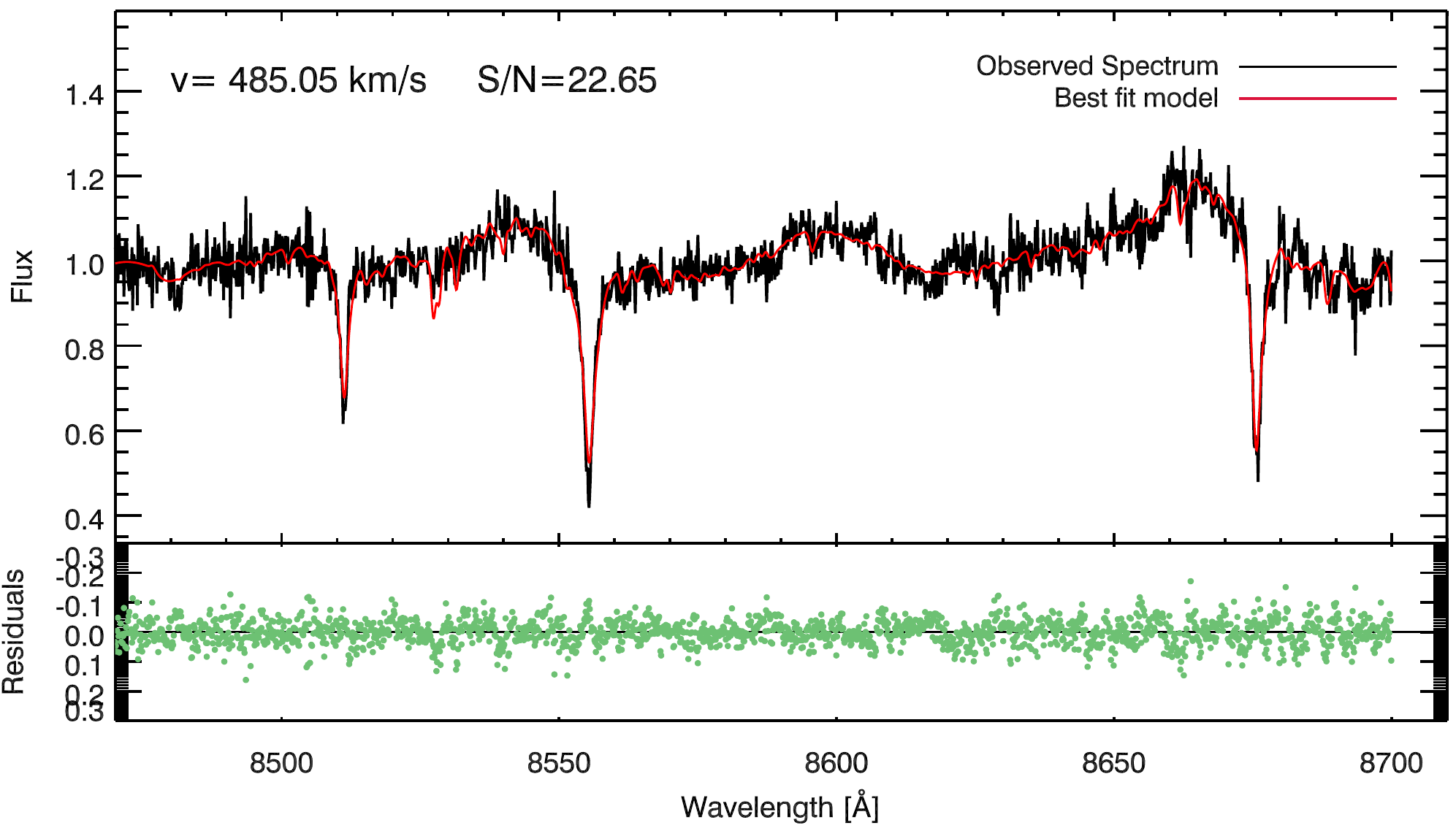}
      \caption{The MIKE spectrum of the calcium triplet region of candidate KV19-442 is shown in black. The best fit derived with pPXF is shown in red and the residuals are shown below in green.} 
         \label{fig:spectra}
   \end{figure}

We find that 5 of our 14 candidates have radial velocities that are consistent with being a member of Centaurus\,A (v$_{\rm helio} = 485-720$~km/s). The other candidates all have $|$v$_{\rm helio}| < 50$~km/s and are thus almost certainly Milky Way foreground stars. 
An example fit zoomed in on the Calcium triplet region of the spectra of KV19-442 is shown in Figure \ref{fig:spectra}. The radial velocities of all confirmed UCDs and the foreground objects are listed in Table \ref{tab:kinematics}. A cut-out image centered on each confirmed UCD is shown in Figure \ref{fig:stamps_confirmed}. 

The four confirmed {\em Gaia} UCD candidates (a-d in Table \ref{tab:kinematics}) were all Rank A (data points marked with a star in Figure \ref{fig:ranking_summary}) showing that the incidence of true UCDs in this ranking category is high. The remaining 9 candidates were of lower ranks (B to D; 5 points marked with a cross in the left panel of Figure \ref{fig:ranking_summary}, whereas 4 of them have no visual assessment and are included in the top histogram of Figure \ref{fig:ranking_summary} with Ranks: twice B, C \& E), consistent with their spectroscopic identification as foreground stars. Additional spectroscopy will be necessary to assess how many UCDs are present in the lower ranks, but it is already clear that both visual classification and colors are useful for effective selection of UCD candidates. 

These newly confirmed UCDs are among the most luminous in Cen\,A. KV19-442 is the 7th and KV19-271 is the 10th most luminous UCDs known in Cen\,A. Fluffy is the most distant known bright globular cluster of Cen\,A in projection, and the clusters KV19-442 and KV19-212 are also more distant than any previously known GC. These globular clusters were missed by previous spectroscopic searches due to being located at such large galactocentric radii. This confirmation of UCDs in such a well-studied nearby galaxy shows both the effectiveness and promise of using \emph{Gaia} to find the best UCD candidates for follow-up spectroscopy.

\begin{deluxetable*}{lccccccccc}
\tablecaption{List of MIKE spectroscopic targets of {\it Gaia}-based UCD candidates \label{tab:kinematics}}
\tablehead{\colhead{Label}  & \colhead{Name}  & \colhead{R.A.} & \colhead{Dec.} & \colhead{$m_{\rm g}$} & \colhead{$M_{\rm g}$} & \colhead{Velocity} & \colhead{S/N} & \colhead{C.L.} & \colhead{Votes}  \\
\colhead{} & \colhead{} & \colhead{} & \colhead{} & \colhead{mag} &  \colhead{mag} & \colhead{km/s} & \colhead{} & \colhead{}} 
\startdata
a & KV19-442 & 202.432394 & -42.391404 & 17.59 & -10.32 & 485.1 $\pm$ 1.2 & 22.7 & 1.71 & 1.0 \\ 	
b & KV19-329 & 201.672201 & -42.703936 & 17.82 & -10.09 & 627.4 $\pm$ 1.6 & 22.3 & 1.67 & 1.0 \\	
c & KV19-271 & 201.296298 & -43.509212 & 17.66 & -10.25 & 548.9 $\pm$ 1.7 & 20.9 & 3.67 & 1.5 \\	
d & KV19-212 & 200.790847 & -43.874459 & 17.93 & -9.98 & 535.1 $\pm$ 2.4 & 21.5 & 2.46 & 1.4 \\  
e & Fluffy &   199.545362	 & -44.157251 &  18.30 & -9.61 & 719.2  $\pm$ 6.4 & 5.2 & - & - \\ 
\hline
- & KV19-054 & 199.1021398 & -44.5545868 & 18.04 & - & 0.5 $\pm$  4.3 & 7.1& 5.08 & - \\  
- & KV19-258 &  201.223936 & -44.931836 & 15.84 & - & -17.3 $\pm$  6.4 & 23.0 & 0.31 & - \\ 
- & KV19-397 &  202.104800 & -42.858302 & 16.26 & -  & -80.8 $\pm$  11.1 & 16.1 & 0.22 & 3.75 \\ 
- & KV19-424  & 202.321732 & -45.212907 & 15.18 & - & -81.3 $\pm$  2.0& 11.6 & $10^{-4}$ & - \\ 
- & KV19-464   & 202.645456  & -44.666926 &17.28 & - & -14.3 $\pm$  32.7 & 16.1 & 0.73 & 1.75 \\ 
- & KV19-492 & 202.862878 & -42.611323 & 17.89 & - &-21.9 $\pm$  6.9  & 8.7 & 0.44 & 2.0 \\ 
- & KV19-521 & 203.050732 & -41.300504 & 17.69 & - & -95.2 $\pm$  29.1 & 10.1 & 2.27 & - \\  
- & KV19-569  & 203.478425 & -42.362577 & 17.89 & - & 23.6 $\pm$  4.6 & 10.8 & 0.30 & 2.0 \\ 
- & KV19-573 &  203.537185 & -42.866562 & 18.06 & - &-19.6 $\pm$  1.7 & 11.7 & 3.35 & 2.0  \\ 
\enddata
\end{deluxetable*}

\section{Conclusion and Application beyond Cen\,A}
\label{sec:discussion}
We use the capabilities of {\em Gaia} to identify new candidate UCDs in the halo of Cen\,A. Our main results are:

\begin{enumerate}

\item 

We used previously-confirmed globular clusters in Cen\,A to show that \emph{Gaia} can be used to identify resolved stripped nuclei and luminous star clusters in nearby galaxies.

\item We derived a relation between the sizes of UCDs, \emph{Gaia} $G$ magnitudes, and the \emph{Gaia} astrometric $BR_{\rm excess}$ parameter. This can be used to estimate sizes accurate to $\sim 30\%$ for extragalactic UCDs with $G<20$ and sizes $\sim 0.1$--0.5\arcsec. We expect this relation can be refined and improved in the future.

\item We apply our new UCD discovery method to obtain a list of {\it Gaia}-based UCD candidates out to 150 kpc from the center of Cen\,A. Of these 632 candidates, 91\% are at larger radii than any previously velocity-confirmed UCD in Cen\,A. Down to the magnitude limit of our search ($G < 19$), our tests suggest that our sample is highly complete, except for the rarest, most extended UCDs.

\item Our \emph{Gaia} UCD sample, while mostly complete, is still affected by foreground and background contaminants. Adding in additional imaging and multi-band photometry helps substantially in ranking the \emph{Gaia}-based sample.

\item We obtained follow-up spectroscopy of 14 UCD candidates, and confirmed all four of the top-ranked sources observed, including two sources which are now in the top 10 of the most luminous UCDs in Cen\,A. 
\end{enumerate}

For future work, we plan to obtain radial velocities and velocity dispersions for all of the good UCD candidates in this paper.
This will enable a large-scale, complete study of stripped nuclei around Cen\,A, which is an important step in fully reconstructing the assembly history of this keystone galaxy.

\subsection{Extension to Other Galaxies}

The \emph{Gaia}-based UCD selection introduced here can be used to find UCDs in a much wider range of galaxies, at least in the distance range $\sim 3$--20 Mpc. The lower limit is approximate, based primarily on the fact that we find that most M31 UCDs/globular clusters (at $D \sim 750$~kpc) are not in \emph{Gaia}, likely because they are too resolved. At the distant end, while a number of Virgo and Fornax UCDs can be found in \emph{Gaia}, the number of sufficiently bright UCDs will decrease substantially at larger distances.

\begin{figure}
   \centering
   \includegraphics[width=\hsize]{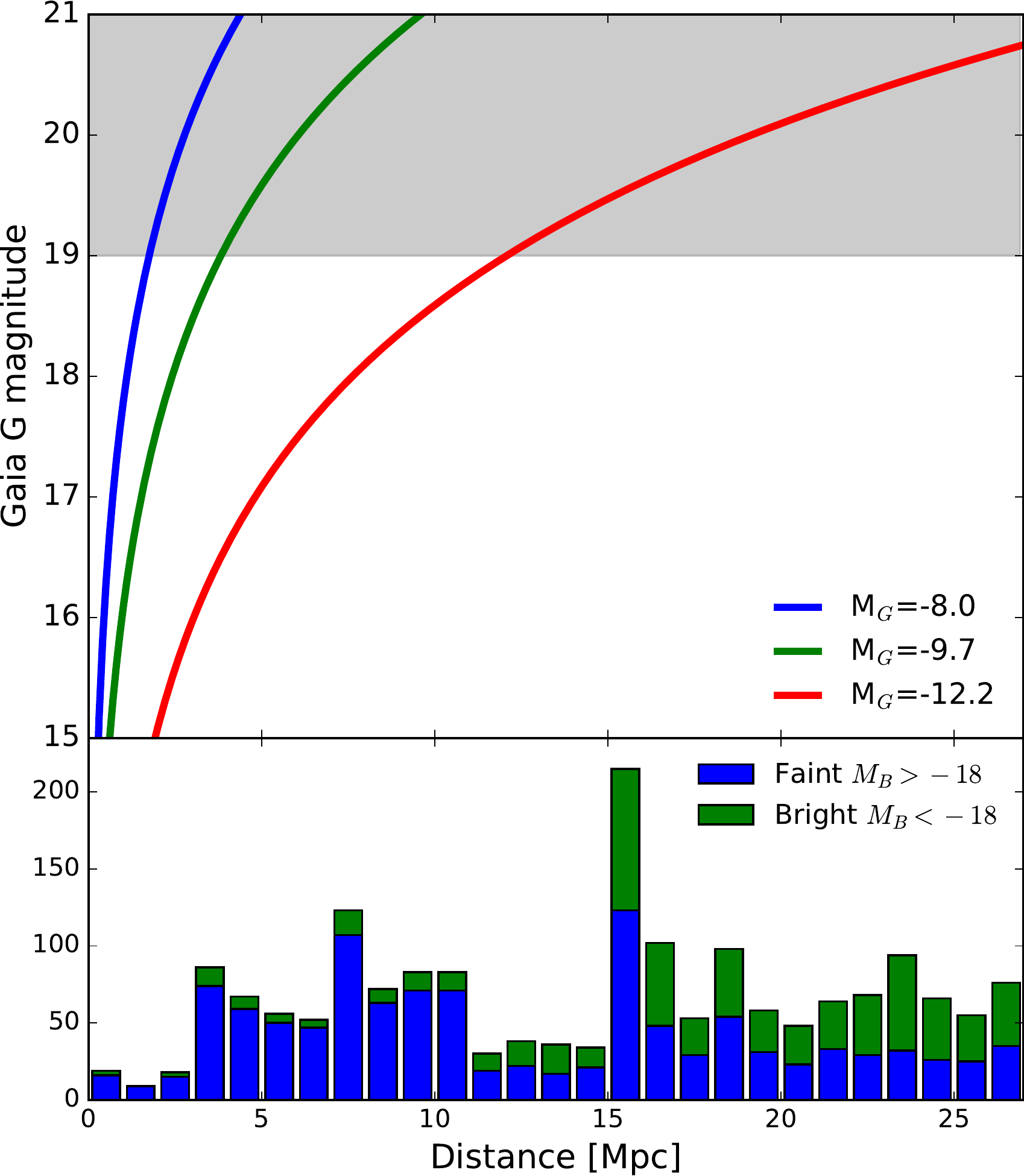}
      \caption{The potential reach of our new {\it Gaia}-based technique for identifying UCDs. The top plot shows how the apparent magnitude of three different fiducial star clusters evolve as function of their distance. The horizontal line marks where the availability of the {\it Gaia} data is essentially complete ($G=19$), with the grey band above ($19 < G < 21$) marking the region where some useful {\it Gaia} information is available.
     The bottom plot is a histogram of the number of galaxies at each distance to show how many bright and fainter galaxies are in principle reachable with this method. For galaxies closer than 11\,Mpc we used the sample of  \cite{Karachentsev2013} and for more distant galaxies the catalog of \cite{Tully2015}.} 
         \label{fig:GCLF}
   \end{figure}

In Figure \ref{fig:GCLF} we show a visual representation of how our UCD selection can be extended to galaxies at different distances and for clusters of different luminosities. The central line is $M_G = -9.7$, equivalent to the selection limit used in the current work to select UCDs with masses above $\sim$10$^6$~M$_\odot$ (where stripped galaxy nuclei are likely to be present; \citealt{Voggel2019}). The rightmost line is $M_G = -12.2$, which corresponds to a mass limit of $\sim10^7$~M$_\odot$, above which
a majority of UCDs appear to be stripped nuclei. The leftmost line is $M_G = -8.0$, roughly the peak of the GC luminosity function at $\sim 2\times10^5$~M$_\odot$. We note that these lines are corrected for the fact that GCs are partially unresolved by Gaia -- for simplicity we use the same 0.52 magnitude factor derived in Section \ref{sec:gaiaphot}; in practice the degree of resolution will likely vary with distance.

We have shown in this paper that at $G < 19$ a nearly complete sample of UCDs can be assembled in Cen A. This same conclusion appears to apply for more distant systems: we find that essentially all of the UCDs with $G \lesssim 19$ in the Virgo UCD catalog of \citet{Liu2015} are in \emph{Gaia}. However, in this same catalog, about half of the UCDs with $19 < G < 21$ are \emph{not} in \emph{Gaia}. Hence down to a fainter limit of $G\sim 20$ to 21 one can construct a reasonable, though not complete, sample of UCDs for more distant galaxies.

The histogram inset to Figure~\ref{fig:GCLF} shows the number of galaxies with $M_{B} < -18$ as a function of distance, where this luminosity limit reflects a guess at those galaxies which could in principle have UCDs detectable with this method. The galaxies are taken from the \cite{Karachentsev2013} and for more distant galaxies from the \cite{Tully2015} catalogue. Even if the sample is restricted to brighter galaxies with $M_{B} < -18$, the number of galaxies for which our method is applicable is appreciable.
In the future we plan to carry out a systematic search for UCDs in the halos of all galaxies in the local Universe where the method is feasible, though spectroscopic follow-up will likely still be necessary.

The discovery potential (in both distance and UCD mass) will likely grow with the upcoming {\em Gaia} data release 3, which will provide improved photometry and astrometry. In any case, the present work is already an exciting step toward a better understanding of luminous globular clusters and stripped nuclei in the local Universe.

\acknowledgments
Work on this project by K.T.V. and A.C.S. was
supported by NSF grants AST-1350389 and AST-1813609. Work on this project by DJS and AH is supported by NSF grants AST-1821967 and 1813708. JS is supposed by NSF grant AST-1812856 and the Packard Foundation. Research by DC is supported by NSF grant AST-1814208, and by NASA through grants number HST-GO-15426.007-A and HST-GO-15332.004-A from the Space Telescope Science Institute, which is operated by AURA, Inc., under NASA contract NAS 5-26555. This research uses services or data provided by the NOAO Data Lab. NOAO is operated by the Association of Universities for Research in Astronomy (AURA), Inc. under a cooperative agreement with the National Science Foundation.

%

\vspace{5mm}
\software{This research made use of Astropy,\footnote{http://www.astropy.org} a 
community-developed core Python package for Astronomy \citep{astropy:2013, astropy2018}. 
This research has made use of NASA's Astrophysics Data System, the Scikit-learn \citep{scikit-learn} code, 
the SciPy \citep{jones_scipy_2001} package and NumPy \citep{van2011numpy}. We also gratefully acknowledge the use of the Carnegie Python Distribution (CarPy) that provided the MIKE data reduction pipeline \citep{Kelson2000, Kelson2003}.}

\facilities{Magellan:Clay (Megacam), {\it Gaia}}





\bibliographystyle{apj}
\bibliography{bibliography_UCDcatalogue}

\appendix

\section{Literature GCs used for completeness assessment}

\startlongtable
\begin{deluxetable*}{cccccccc}

\tablecaption{List of the 57 previously confirmed UCDs that were used to assess completeness in Figure 1 \label{tab:confirmed_ucds}. The Taylor 2017 g' magnitudes are from their Table 2 and are already extinction corrected. \label{tab:lit_gcs}}
\tablehead{\colhead{Taylor 2017 Name}  & \colhead{R.A.}  & \colhead{DEC.} & \colhead{Gaia G} & \colhead{AEN} & \colhead{$BR_{\rm excess}$} & \colhead{Taylor 2017 g'}  \\
\colhead{} & \colhead{$^{\circ}$} & \colhead{$^{\circ}$} & \colhead{} & \colhead{}& \colhead{} &  \colhead{mag}} 
\startdata
T17-1002 & 200.909654 & -42.773053 & 18.87 & 9.13 & 2.40 & 18.09 \\
T17-1008 & 200.926390 & -43.160493 & 19.54 & 15.29 & 2.82 & 18.76 \\
T17-1020 & 200.956660 & -43.242243 & 19.09 & 6.73 & 2.10 & 18.62 \\
T17-1050 & 200.994829 & -43.026381 & 19.31 & 10.01 & 2.75 & 18.70 \\
T17-1110 & 201.075206 & -42.816957 & 18.91 & 6.53 & 2.39 & 18.16 \\
T17-1188 & 201.153561 & -43.321188 & 19.42 & 9.03 & 2.31 & 18.73 \\
T17-1197 & 201.162357 & -43.335133 & 18.72 & 3.86 & 2.09 & 18.38 \\
T17-1202 & 201.168229 & -43.301482 & 18.46 & 6.36 & 2.46 & 17.93 \\
T17-1203 & 201.168337 & -43.584709 & 19.44 & 7.22 & 2.50 & 18.73 \\
T17-1207 & 201.170003 & -42.683779 & 18.69 & 2.71 & 1.91 & 18.43 \\
T17-1232 & 201.194548 & -43.021799 & 19.43 & 8.58 & 3.10 & 18.74 \\
T17-1243 & 201.200123 & -43.137294 & 18.82 & 4.90 & 2.56 & 18.30 \\
T17-1260 & 201.211852 & -43.023042 & 19.13 & 5.19 & 2.98 & 18.77 \\
T17-1264 & 201.214461 & -43.203099 & 18.86 & 4.47 & 2.58 & 18.24 \\
T17-1284 & 201.226439 & -42.890201 & 17.39 & 3.29 & 2.03 & 17.10 \\
T17-1287 & 201.227938 & -43.022712 & 18.20 & 5.12 & 2.74 & 17.78 \\
T17-1313 & 201.239246 & -43.018940 & 18.47 & 4.11 & 2.49 & 18.27 \\
T17-1314 & 201.239568 & -42.989816 & 19.13 & 7.50 & 3.06 & 18.58 \\
T17-1322 & 201.242499 & -42.936123 & 17.98 & 3.03 & 2.47 & 17.60 \\
T17-1347 & 201.254756 & -42.947666 & 18.94 & 4.20 & 2.42 & 18.62 \\
T17-1358 & 201.257506 & -43.157080 & 18.58 & 10.72 & 3.02 & 17.82 \\
T17-1375 & 201.263984 & -42.846132 & 18.64 & 3.58 & 2.33 & 18.32 \\
T17-1386 & 201.269934 & -43.160808 & 18.99 & 9.41 & 2.77 & 18.43 \\
T17-1388 & 201.270881 & -42.954239 & 17.53 & 3.96 & 2.15 & 17.27 \\
T17-1395 & 201.273696 & -43.175240 & 18.24 & 3.95 & 2.52 & 17.85 \\
T17-1400 & 201.275932 & -43.253251 & 19.12 & 8.39 & 2.78 & 18.52 \\
T17-1430 & 201.292744 & -42.892504 & 18.16 & 4.91 & 2.65 & 17.65 \\
T17-1432 & 201.293668 & -42.747977 & 18.30 & 5.61 & 2.37 & 17.72 \\
T17-1447 & 201.300810 & -43.276086 & 19.30 & 6.39 & 2.29 & 18.72 \\
T17-1455 & 201.303631 & -43.133121 & 18.27 & 3.41 & 2.16 & 18.06 \\
T17-1486 & 201.317717 & -42.848126 & 18.91 & 8.17 & 2.58 & 18.31 \\
T17-1596 & 201.376623 & -43.197138 & 19.05 & 13.03 & 3.02 & 18.46 \\
T17-1611 & 201.382221 & -43.322982 & 19.08 & 8.44 & 2.36 & 18.48 \\
T17-1616 & 201.386606 & -43.117305 & 18.95 & 4.60 & 2.66 & 18.63 \\
T17-1639 & 201.395821 & -42.601374 & 18.90 & 5.17 & 2.35 & 18.41 \\
T17-1664 & 201.407701 & -42.941120 & 18.87 & 3.08 & 2.85 & 18.76 \\
T17-1682 & 201.415494 & -42.933094 & 18.04 & 8.38 & 2.86 & 17.48 \\
T17-1688 & 201.419139 & -43.353852 & 18.68 & 4.35 & 2.36 & 18.26 \\
T17-1719 & 201.430767 & -43.123036 & 18.60 & 9.20 & 2.93 & 18.01 \\
T17-1773 & 201.457002 & -42.913691 & 18.31 & 6.30 & 2.72 & 17.83 \\
T17-1780 & 201.458071 & -42.869243 & 19.07 & 5.22 & 2.69 & 18.58 \\
T17-1805 & 201.469692 & -43.096254 & 18.20 & 4.98 & 2.51 & 17.74 \\
T17-1814 & 201.473123 & -42.985426 & 18.37 & 4.49 & 2.64 & 18.01 \\
T17-1887 & 201.505241 & -43.570995 & 18.83 & 5.73 & 2.46 & 18.26 \\
T17-1899 & 201.511786 & -42.949167 & 18.81 & 4.74 & 3.05 & 18.46 \\
T17-1920 & 201.522473 & -42.942327 & 17.56 & 5.45 & 2.54 & 17.02 \\
T17-1940 & 201.532144 & -42.866721 & 18.67 & 4.95 & 2.33 & 18.26 \\
T17-1957 & 201.544021 & -42.895178 & 18.53 & 3.20 & 2.15 & 18.37 \\
T17-1973 & 201.563543 & -42.808167 & 18.43 & 5.518 & 2.57 & 18.18 \\
T17-1974 & 201.566080 & -42.916914 & 18.12 & 1.76 & 1.78 & 17.89 \\
T17-1989 & 201.581834 & -43.055186 & 19.15 & 6.43 & 2.63 & 18.65 \\
T17-2013 & 201.599019 & -42.900293 & 18.69 & 5.35 & 2.27 & 18.37 \\
T17-2018 & 201.600676 & -42.783526 & 18.84 & 6.78 & 2.37 & 18.41 \\
T17-2064 & 201.660367 & -42.762706 & 19.08 & 8.37 & 2.61 & 18.50 \\
T17-2091 & 201.689029 & -43.442810 & 19.06 & 4.47 & 2.13 & 18.59 \\
T17-2107 & 201.724630 & -43.321591 & 18.50 & 3.90 & 2.02 & 18.19 \\
T17-2133 & 201.764179 & -42.454757 & 18.66 & 3.99 & 2.27 & 18.27 \\
\enddata
\end{deluxetable*}



\end{document}